\documentclass{article}
\usepackage{amssymb}
\usepackage{amsbsy}
\usepackage{enumerate}
\usepackage{graphicx}
\usepackage{tikz}
\usepackage{datetime}

\newcommand{\be}{\begin{equation}}
\newcommand{\ee}{\end{equation}}

\def\bea{\begin{eqnarray}}
\def\eea{\end{eqnarray}}
\def\bean{\begin{eqnarray*}}
\def\eean{\end{eqnarray*}}

\newcommand{\barr}{\begin{array}}
\newcommand{\earr}{\end{array}}

\newcommand{\bed}{\begin{displaymath}}
\newcommand{\eed}{\end{displaymath}}
\newcommand{\bal}{\begin{array}{ll}}
\newcommand{\eal}{\end{array}}

\def\mc#1{\mathcal#1}

\begin{document}


\centerline{\Large\bf  Nambu and the Ising Model}
\vskip .5cm
\centerline{Lars Brink}
\centerline{\it Department of Physics, Chalmers Institute of Technology}
\centerline{\it S-41216 G\"oteborg, Sweden}
\vskip .2cm\centerline{Pierre Ramond}
\centerline{\it University of Florida}
\centerline{\it Gainesville Florida 32611, United States}
\vskip .5cm

\section{Abstract}
2021 was Y\^oichir\^o Nambu's birth centenary. To mark the occasion, we engaged in writing a historical/scientific description of his most incisive papers, to be published by WSP. This turned out to be a most demanding but also rewarding enterprise. Most papers we have chosen are world classics, but  Nambu was the humblest genius we have known, and we expected to find some of his greatest unknown insights. 

One was very early in his career;  on September 1, 1949,  the physics journal founded by H. Yukawa
  {\em Progress of Theoretical Physics} 
 received\footnote{Y. Nambu, { Prog.  Theo. Phys. {\bf 5}, 1, (1950)}},
\vskip .2cm
\centerline {\em ``A Note on the Eigenvalue Problem in Crystal Statistics".} 
\vskip .2cm
\noindent Written in  Osaka in the newly started group created for him, it contained a curious acknowledgment  {\em ``The main part of the present work had been completed  nearly two years ago. It is through the kindness of Professor Husimi and Mr. Sy\^ozi of Osaka University that the author enjoys the opportunity of publishing this note."}\footnote{News of  the Lamb-Retherford experiment reached Japan in the September 29 1947 issue of  {\em Time Magazine}. }    We think it was  at the suggestion of Professor  Sin-Itiro Tomonaga, that Nambu  put aside  his crystal work to calculate the Lamb shift, which he published independently but after Schwinger.
\vskip .2cm

We find this paper and the techniques developed in it so remarkable and even relevant for today that we have decided to write this  in greater detail than the one in the book.

\section{Introduction}
 Y{\^o}ichir{\^o} Nambu came back to Tokyo University after the war. Nambu was always a very modest person never boasting about his achievements, but it is clear that he must have been an outstanding undergraduate,  graduating in 2.5 years since it was cut short by half a year. Upon graduation he was awarded  a graduate fellowship. He came back to a devastated Tokyo and for three years he lived in his office sleeping on his desk mainly living on potatoes. It is difficult for us to understand how hard life was, the years after the war. 

The older theoretical physicists at Tokyo University were mostly engaged in research in statistical physics and condensed matter theory, and it was natural for Nambu to take up a study of  the 2-dimensional  Ising model which had been solved by Lars Onsager in 1944. He was well into a new formulation and solution of the model in the fall of 1947 when news reached Japan about the experimental discovery of the Lamb shift. When asked by Tomonaga to study this problem, he put his work on the Ising model aside.  Two years later he took it up again and published the result. 

The paper is Nambu's  formulation of the  two-dimensional Ising model. Compared with Onsager's formidable solution\footnote{L. Onsager, Phys. Rev. {\bf 65}, (1944) 117} which diagonalized a $(2^N\times 2^N)$, Nambu's Ising model lives in a $2N$-dimensional Hilbert space of $N$ qubits. A  four-page computation of the eigenvalues of the transfer matrix suffices to  reproduce Onsager's results! It is  a remarkable unexplored  aspect of this well-studied system. Nambu said at much, in his characteristically humble manner: 
\vskip .3cm
 \noindent {\it Though as yet no substantial applications has been attempted, nor anything physically new has been derived, it may be hoped that it will do some profit for those who are interested in such problems.} 
\vskip .2cm
A little background on the Ising model and its place in  physics: when  asked after the war if anything new had happened in fundamental physics, Pauli replied ``not much, except for Onsager's solution of the Ising Model". 
\vskip .3cm
In 1920, Wilhelm Lenz   suggested   {\it ``Beitrag zum Verst\"andnis der magnetischen Erscheinungen in festen K\"orpern"}\footnote{W. Lenz, Physik. Z. XXI, 1920, 613-615}  
 that ferromagnetism could be explained in terms of interacting nearest-neighbor magnets which could  flip in opposite directions (``umklapping"). He asked his student Ernst Ising to solve his model. Ising  did find an analytical solution\footnote{E. Ising, Zeits. f. Physik 31, 253 (1925)}  
{\it ``Beitrag zur Theorie des Ferromagnetismus"} but only on a linear lattice and found no ferromagnetic transition.

\section{Nambu's Crystal Statistics}
 In 1944 came Lars Onsager's epochal analytic solution which inspired Nambu's paper we now present 

\vskip .3cm
\noindent {\bf The  Linear Single Spin Array}
\vskip .2cm
\noindent Nambu first discussed  the simplest one-dimensional array of $N$ identical particles  with a different  two-valued spin at each site, $n=1,2,...,N$,  with  nearest-neighbor interactions,

 $$ P=\sum_{n=1}^NP^{}_{n,n+1}=\sum_{n=1}^N\frac{1+\sigma_n\,\sigma_{n+1}}{2}.$$ 
 Inspired by quantum field theory, he introduced a different fermi oscillators at each site, and commute with one another at different  sites,

$$\{a_n^{},a^\dagger_n\}=1,\qquad [a_n^{},a^\dagger_m]=0,~~n\neq m,$$
with periodic boundary conditions. In terms of these,  
$$ 
P=\sum_{n=1}^N\,[a^\dagger_n a^{}_{n+1}+a^\dagger_{n+1}a^{}_n+2a^\dagger_n a^{}_na^\dagger_{n+1} a^{}_{n+1}-2a^{\dagger}_n a^{}_n +1].$$
This is the conventional approach.  Now comes Nambu's  fundamental observation: $P$, as a function of quadratic combinations,  is the same whether  the operators at different sites  commute or anticommute. In an audacious leap, Nambu  suggested  an alternate description of $P$ in terms of new ladder operators 
$$ \{a_n^{},a_m^{}\}=\{a_n^{\dagger },a_m^\dagger\}=0 ;\qquad \{a_n^{},a_m^{\dagger}\}=\delta^{}_{m,n},$$ 
for all $n,m$. $P$ now lives in a much smaller $2N$-dimensional Hilbert space, rather than in a $2^N$-dimensional one in the conventional approach. It should lead to the same physics. 

The rest of the paper is the exploitation of this generalization, first for $P$, then for the isotropic $X-Y$ model, and culminating in a much simpler solution of the Ising model for both square and ``screw" arrays. 
\vskip .3cm
 
Nambu wrote $P$ in a manifestly Hermitian form,
$$ 
P=\sum_{n,m}^N\Big(a^\dagger_na^{}_{m}\delta(n-m+1)-a_{n}^\dagger a_m^{}\delta(n-m) +a^\dagger_na^{}_n\,a^\dagger_{m}a^{}_{m}\delta(n-m+1)+~{\rm h.c.}~\Big),
$$ 
and introduced the operator Fourier transforms,

$$ \tilde a^{}_k=\frac{1}{\sqrt{N}}\sum_na^{}_n\eta^{k},\quad \tilde a^{\dagger}_k=\frac{1}{\sqrt{N}}\sum_na^{\dagger}_n\eta^{-k}.
$$ 
 where $\eta$  are the $N$ roots of unity ($\eta=e_{}^{\frac{2\pi i}N}$). $P$ emerges as,  
 
 $$ 
 P=\sum_{k=1}^N\big[ \tilde a^\dagger_k\tilde a^{}_k\eta^{-k}-\tilde a^\dagger_k\tilde a^{}_k+c.c.\big]+\sum_{k,l=1}^N
 \tilde a^\dagger_k\tilde a^{}_k\tilde a^\dagger_l\tilde a^{}_l\delta(k-l\pm 1).
 $$ 
Nambu interpreted $P$ as  Hamiltonian sum of a quadratic ``kinetic" term and  a quartic expression describing a  ``short-range" (across the sites)  potential. 

For  large $N$, the potential becomes insignificant and $P$ describes  ``free"  $N$ Bloch spin waves with energies $\epsilon_k=2(\cos\frac{2\pi}{N}k-1),\, k=1,2, \dots, N$. He concluded that it was   {\it ``... a good approximation when the magnetization is nearly complete (low temperature)"}.

Nambu  now  applies his formalism to the isotropic $X-Y$ model, where he introduces new techniques before considering the Ising model.

\vskip .3cm
 \noindent{ \bf The Isotropic X-Y Model}
\vskip .2cm
\noindent In order to simplify the notation, Nambu replaced the ladder operators   by  $2N$ ``real  coordinates" $(a^\dagger_n+a^{}_n),\,i(a^\dagger_n-a^{}_n)$,  that is Grassmann coordinate and momentum at each site. They  span an orthogonal basis in a $2N-$dimensional vector space,
 $$  \{x^{}_n,\,x^{}_m\}=2\delta^{}_{rs},\quad  n,m=1,2,\dots 2N.$$ 
 The permutation operator of the ``Isotropic X-Y" model includes two Pauli spin matrices, $\sigma_x$ and $\sigma_y$, 

$$ P_{X-Y}=\sum_{n=1}^{N}\,(\sigma^{}_{n,x}\sigma^{}_{n+1,x}+\sigma^{}_{n,y}\sigma^{}_{n+1,y})\equiv \sum_{n=1}^{N}\,(A^{}_n+B^{}_n).$$ 
The new operators $A_n$ and $B_n$ commute, except at  adjacent sites where they anticommute, 
$$ \{A^{}_n,B^{}_{n\pm1}\}=0,$$ 
 and obey the constraints $  A^2_n=B^2_{n}=1$. These algebraic requirements are solved by expressing $A_n$ and $B_n$ as quadratic  combinations of  ``Nambu's basis" coordinates $\{x_n\}$, 

$$ A^{}_n=ix^{}_{2n}x^{}_{2n+1},\quad B^{}_n=ix^{}_{2n-1}x^{}_{2n+2},\qquad n=1,2,\dots, N,$$ 
so that $A_n$ links adjacent sites and $B_n$ hops over three sites. For $N$ even, and  periodicity, 
the   constraints collapse into one $1=\prod A^{}_n=\prod B^{}_n\equiv x$. 
  
As in  the one-spin linear case, Nambu introduced Fourier transforms

$$  
\tilde x^{}_k\equiv \frac{1}{\sqrt{2N}}\sum_{n=1}^N x^{}_{2n}\eta_{}^{nk},
\qquad 
\tilde y^{}_k\equiv  \frac{1}{\sqrt{2N}}\sum_{n=1}^N x^{}_{2n+1}\eta_{}^{nk},
 $$
for  even an odd sites. They describe for each $k$ two fermion oscillators since, 

$$ \{\tilde x^{}_k,\tilde x^{}_{-l}\}=\{\tilde y^{}_k,\tilde y^{}_{-l}\}=\delta^{}_{kl},\quad \{\tilde x^{}_k,\tilde y^{}_{-l}\}=0,$$ 
 where $k,l$ run from $-N$ to $N$ in integer steps. Then
 
$$ 
 P^{}_{X-Y}=-2\sum_{k=1}^{N}\big(\tilde x^{}_k\tilde y^{}_{-k}+\tilde y^{}_k\tilde x^{}_{-k}\eta^{2k}_{}\big).
$$ 
After some algebra,  

$$
 P^{}_{X-Y}= -4\sum_{k=1}^{N/2}z^{}_k\sin\frac{2\pi k}{N},
$$ 
 where
 
 $$z^{}_k\equiv \tilde x^{}_k\tilde y^{}_{-k}\eta_{}^{-k}-\tilde x^{}_{-k}\tilde y^{}_k\eta_{}^k.$$
 is a sum of  quadratic forms in $\tilde x_k$ and $\tilde y_l$, which is readily be diagonalized. 
  \vskip .3cm
  \noindent For each $k$, Nambu found the  $(4\times4)$ matrix representation,
  
   \bean
 \tilde x^{}_k&=&\sigma^{}_-\otimes \sigma^{}_3,\qquad \tilde x^{}_{-k}=\sigma^{}_+\otimes \sigma^{}_3,
 \\\tilde y^{}_k&=&\sigma^{}_0\otimes \sigma^{}_-,\qquad \tilde y^{}_{-k}=\sigma^{}_0\otimes \sigma^{}_+,
\eean
that is,
 
 $$ 
z^{}_k=\begin{pmatrix}{0&0&0&0\cr
0&0&-e_{}^{\frac{2\pi ik}{N}}&0\cr
0&-e_{}^{-\frac{2\pi ik}{N}}&0&0\cr
0&0&0&0}\end{pmatrix}.
$$ 
The eigenvalues follow,

$$ \epsilon _k^{}= 0,0,1,-1,~~\longrightarrow~~P^{}_{X-Y}=4\sum_{k=1}^{N/2}\,\epsilon^{}_k\sin\frac{2\pi k}{N},
$$ 
but restricted by the one boundary condition  $x=1$,  ($N$ even).  $x$ can be expressed in terms of rotations,  $$
 R(\theta)\equiv e_{}^{2\theta\sum_n x^{}_{2n}x^{}_{2n+1}}= \prod_{n=1}^{N/2}e^{2\theta (\tilde x^{}_k\tilde y^{}_{-k}+\tilde x^{}_{-k}\tilde y^{}_{k})}. $$
  It  can be expressed as,
 
 $$ 
R(\theta)\equiv \prod_{n=1}^{N/2}e^{2\theta (\tilde x^{}_k\tilde y^{}_{-k}+\tilde x^{}_{-k}\tilde y^{}_{k})}
 _{}= \prod_{n=1}^{N/2}\big[1+(\cos(2\theta)-1)R^{2}_k+i\sin(2\theta)R^{}_k\big],
 $$
 where 
 $$ 
 R^{}_k\equiv i(\tilde x^{}_k\tilde y^{}_{-k}+\tilde x^{}_{-k}\tilde y^{}_{k}),
 $$
  commutes with $z_k$ and satisfies $ R^{}_k=R^3_k$. Comparing these two expressions at  $\theta=\pi/2$, 
 
 $$R(\pi/2)=x_2^{}x_4^{}\dots x^{}_{2N}x^{}_1=-x _{}= \prod_{k=1}^{N/2}(1-2R^2_k).$$
 
 $$
 R(\theta)\equiv \prod_{n=1}^Ne^{\theta x_{2n}^{}x^{}_{2n+1}},\quad R(\pi/2)=x_2^{}x_4^{}\dots x^{}_{2N}x^{}_1=-x
 _{}.
 $$ 
 The number of non-zero eigenvalues is restricted to 
 
 $$ x=1~~\longrightarrow ~~~~\prod_{n=1}^{N/2}(1-2\epsilon_k^2) =(-1)^{N/2}_{}.$$
which completes the solution of the isotropic $X-Y$ model.  The next sections will truly highlight the power of his method applied to the two-dimensional Ising model.

\vskip .3cm
\noindent{\bf The Square Ising Model}
\vskip .2cm
\noindent Nambu's starting point is  Onsager's operator (neglecting the prefactor) which describes the square Ising model with different interaction strengths for vertical and horizontal nearest neighbors $J$ and $J'$,

$$  \mc H=\exp{ \big[H'\sum_{n=1}^N s^{}_ns^{}_{n+1}\big]}\,\exp{ \Big[H^*_{}\sum_{n}^N c^{}_n\Big]},$$ 
 where $H'=J'/kT$, and  $ H^*$  is the the Kramers-Wannier\footnote{H.A. Kramers and G. H. Wannier, Phys. Rev. {\bf 60} (1941) 252}  dual of $H=J/kT$. The spins satisfy,
$$ s_n^2=c^2_n=1, \qquad \{\,s_s,\,c_n\}=0,$$ 
and commute  with one another at different sites.  
\vskip .2cm
Onsager's {\it tour de force} was  to determine the eigenvalues of this operator, and prove the existence of a ferromagnetic transition in the thermodynamic limit.
\vskip .3cm

\noindent 
As he did for the  $X-Y$ model, Nambu introduced new variables,
$$ 
S^{}_n\equiv s^{}_ns^{}_{n+1},\qquad C^{}_n\equiv c^{}_n,$$ 
which commute with one another except at adjacent sites,
$$ 
\{\,S^{}_n,\,C^{}_{n\pm 1}\,\}=0.$$
and  boundary conditions, 
$ S\equiv S_1^{}S_2^{}\cdots S_N=1,\,\, C\equiv C_1^{}C_2^{}\cdots C_N=\pm 1.$ 

\vskip .3cm
\noindent   $S^{}_n$ and $C^{}_n$ are now expressed in the ``Nambu basis"  $\{x\}$,

\be 
S^{}_n=ix^{}_{2n}x^{}_{2n+1},\quad C^{}_n=ix^{}_{2n-1}x^{}_{2n},\ee 
for even $N$ with boundary conditions,  

$$
C=i^Nx^{}_1x^{}_2x^{}_3x^{}_4\cdots x^{}_{2N-1}x^{}_{2N}\equiv X,\quad
S=i^Nx^{}_2x^{}_4x^{}_2x^{}_5\cdots x^{}_{2N}x^{}_1=-X.$$ 

The stage is set for Nambu's computation of the eigenvalues and eigenfunctions of the transfer matrix,

\be 
\mc H=\exp{\Big[iH'\sum x^{}_{2n}x^{}_{2n+1}\Big]}\,\exp{\Big[i H^*_{}\sum x^{}_{2n-1}x^{}_{2n}\Big]}\equiv \mc H_2\mc H_1.\ee 
in the  $\{x\}$ basis.
\vskip .3cm
\noindent It is a product of operators, 
$$
\mc U=e_{}^{\,\theta x^{}_nx^{}_m}, ~~n\neq m,\quad 
e_{}^{{\theta}/{2}\, x^{}_n\,x^{}_m}=\cos\theta+\sin \theta\,x^{}_nx^{}_m,$$ 
which describe a rotation by $\theta$ in the $x_n-x_m$ plane.
 $\mc H$  is just a rotation by an angle $iH^*_{}$ followed by another rotation by $iH'$. In some basis $\{x'\}$,  $\mc H$ will be expressed in Jordan's canonical form, 
 $$
\mc H=\exp{\Big[i\sum x'_{2n}x'_{2n+1}\gamma^{}_n\Big]}$$
with eigenfunctions that satisfy $ \mc H\Psi=e^{\sum\epsilon_n\gamma^{}_n}_{}\Psi,\quad  
 \epsilon_n=\pm 1$, the largest eigenfunction is simply $ \mc H^{}_{\rm max}=e^{\sum |\gamma_n|}_{}.$   
\vskip .3cm
\noindent Using periodicity, $x^{}_1=x_{2n+1}$,  we obtain,
\bean \mc H_1&=&\exp{\Big[iH^*(x_1x_2+x_3x_4+\cdots x_{2N-1}x_{2N})\Big]},\\
 \mc H_2&=&\exp{\Big[iH'(x_2x_3+x_4x_5+\cdots x_{2N}x_{1})\Big]},
\eean
 so that  $\mc H_1$ rotates the ``odd-even" pairs $(x^{}_{2n-1},\,x^{}_{2n})$, 
\be
\mc H_1: ~~ \begin{pmatrix}{x^{}_{2n-1}\cr x^{}_{2n}}\end{pmatrix}~~\longrightarrow~~\begin{pmatrix}{y^{}_{2n-1}\cr y^{}_{2n}}\end{pmatrix}=\mc R(2iH^*) 
\begin{pmatrix}{x^{}_{2n-1}\cr x^{}_{2n}}\end{pmatrix},
\ee
while $\mc H_2$ rotates the  ``even-odd" pairs $(y^{}_{2n},\,y^{}_{2n+1})$,

\be
\mc H_2: ~~ \begin{pmatrix}{y^{}_{2n}\cr y^{}_{2n+1}}\end{pmatrix}~~\longrightarrow~~
\begin{pmatrix}{z^{}_{2n}\cr z^{}_{2n+1}}\end{pmatrix}=\mc R(2iH')
\begin{pmatrix}{y^{}_{2n}\cr y^{}_{2n+1}}\end{pmatrix}
\ee
where
$$\mc R(2it)=\begin{pmatrix}{\cos(2it)&\sin(2it)\cr -\sin(2it)&\cos(2it)}\end{pmatrix}
=\begin{pmatrix}{\cosh(2t)&i\sinh(2t)\cr -i\sinh(2t)&\cosh(2t)}\end{pmatrix}
$$
The combined action of  $\mc H=\mc H_2\mc H_1$  amounts to a linear transformation on the original pair,

\be 
\begin{pmatrix}{x^{}_{2n-1}\cr x^{}_{2n}}\end{pmatrix}~~~\longrightarrow~~~ \begin{pmatrix}{z^{}_{2n-1}\cr z^{}_{2n}}\end{pmatrix}\equiv \lambda\begin{pmatrix}{x^{}_{2n-1}\cr x^{}_{2n}}\end{pmatrix},
\ee 
where $\lambda$ is the eigenvalue. Define  the coefficients 

$$ a=i\cosh(2H^*)\sinh(2H'),\quad b=i\sinh(2H^*)\cosh(2H')$$ 
$$ c= -\sinh(2H^*)\sinh(2H'),\quad d=\cosh(2H^*)\cosh(2H')$$
with
$$ ab-cd=0,\quad a^2+b^2+c^2+d^2=1.$$
\vskip .2cm
Nambu's clever choice of the pairs on which $\mc H$ acts, has reduced the characteristic equation to two equations for each $n$, 
\bean 
-bx^{}_{2n-1}+(d-\lambda)x^{}_{2n}+ax^{}_{2n+1}+cx^{}_{2n+2}&=&0,\\
-ax^{}_{2n}+(d-\lambda)x^{}_{2n+1}+bx^{}_{2n+2}+cx^{}_{2n-1}&=&0.
\eean 
Nambu's elegant solution of these equations is  to introduce two matrices and an eigenfunction,

$$ A=\begin{pmatrix}{a&c\cr \lambda-d&-b}\end{pmatrix},\quad B=\begin{pmatrix}{b&\lambda-d\cr c&-a}\end{pmatrix},\quad  \psi^{}_n=\begin{pmatrix}{x^{}_{2n-1}\cr x^{}_{2n}}\end{pmatrix},$$ 
 so that the  characteristic  equations become one matrix equation, 
 
$$ A\psi^{}_{n+1}=B\psi^{}_{n},$$ 
resulting in a recursion relation ($A$ is not singular),
$$
~\psi^{}_{n+1}=A^{-1}_{}B\psi^{}_{n}\equiv D\psi^{}_n,
$$
that is readily solved,
$$ \psi^{}_{n+1}=D^n_{}\psi^{}_1.$$ 
Finally, the periodicity constraint  $\psi_{N+1}=\psi_1$ leads to the characteristic equation,
\be \det(1-D^N)=0.\ee 
It is solved by means of the ``well-known" identity,
\be 
1-D^N=\prod_{k=1}^N(\eta^k_{}-D),\qquad \eta=e_{}^{\frac{2\pi i}{N}},\ee 
which reduces to $N$ equations,
$$ \det( \eta_{}^k -D)=0~~~\longrightarrow~~~|A\eta^k_{}-B\,|=0.\quad k=1,2\dots, N. $$ 
Explicitly,

$$  \Big |
\begin{matrix}{\eta^k_{}a-b&\eta^k_{}c-(\lambda-d)\cr \eta^k_{}(\lambda-d)-c&-\eta^k_{}b+a}\end{matrix}\Big |=0.
$$ 
This simple quadratic equation,  

$$ \lambda^2-2\lambda\big[d +c\cos\varphi^{}_k\big]+1=0,\qquad \varphi_k= \frac{2\pi k}{N},$$
has two solutions for each $k$,

\be \lambda_{k\pm}^{}=\cosh(2\gamma_k)\pm \sinh(2\gamma_k),\ee
with
\be
\cosh2\gamma^{}_k=d+c\cos\varphi_k=\cosh(2H^*)\cosh(2H')-\sinh(2H^*)\sinh(2H')\cos\varphi_k,
\ee
 the same formula as Onsager's Eq(95) of his paper:
$$ 
\frac{1}{2}\sum_{r=1}^n \gamma_{2r-1}=\frac{1}{2}\sum_{r=1}^n\cosh^{-1}\big[\cosh2H'\cosh 2H^*-\sinh2H'\sinh2H^*\cos((2r-1)\pi/2n))],
$$
with largest eigenvalue, 

$$ H^{}_{\rm max}=\exp\big[\sum_k|\gamma_k|\big]. 
$$
By a simple series of steps, Nambu duplicated Onsager's result! It is a conceptual result, the Ising model realized from a $2N$-dimensional Hilbert space.
\vskip .2cm
Nambu also pointed out that this method applies {\it mutatis mutandis}  (when necessary changes made) to certain variants of Onsager model such as the honeycomb lattice of Kodi Husimi and Itiro Sy\^ozi\footnote{K. Husimi and I. Sy\^ozi, Prog. Theo. Phys. V, (1950) 177} . 

When he  tried to apply his method to the three-dimensional case in his basis, he found that not all operators are  exponentials  of quadratics (i. e. rotations), some are exponentials of quartics, such as $e^{ax_1x_2x_3x_4}_{}$. In view of Nambu's many prescient comments, it might be interesting to follow his path,  although  no analytic solution has ever been found.

\vskip .3cm
\noindent{\bf The Helical Ising Model} 
\vskip .2cm
\noindent  In their attempt to find an analytic solution for Ising's model, Kramers and Wannier argued in $1941$ that it was simpler  to describe the lattice in terms of one string of spins, lying on the the wires of an infinite solenoid, which they call the {\em ``screw lattice"}.  Nambu noted that {\em ``This model seems more convenient for general purposes than that used by Onsager."}
 \vskip .3cm
 \noindent To transform Onsager's expression into the Kramers-Wannier helical string model,  Nambu rearranged the interaction as
 
 \bean
 \mc H=\mc H_1\mc H_2&=&e_{}^{iH^* x_1x_2}\,e_{}^{iH^* x_3x_4}\cdots \,e_{}^{iH' x_2x_3}\,e_{}^{iH' x_4x_5} \cdots\\
  &=&e_{}^{iH^* x_1x_2}\,(e_{}^{iH^* x_3x_4} \,e_{}^{iH' x_2x_3})(e_{}^{iH^* x_5x_6} \,e_{}^{iH' x_4x_5})\cdots ,\\
  &=&e_{}^{iH^* x_1x_2}\prod_{n=1}^{N}\,\mc H^{}_ne_{}^{iH' x_{2N}x_1},\qquad  \mc H^{}_n=e_{}^{iH^* x_{2n+1}x_{2n+2}} \,e_{}^{iH^* x_{2n}x_{2n+1}}.\eean
 Neglecting the two boundary terms, he started from,
 
 \be
 \mc H=\prod_{n=1}^{N} \mc H^{}_n\ee
 \vskip .3cm
 \noindent 
 First step is to express  a  displacement operator $P$  as a product of rotations,
$$x^{}_n~~~\longrightarrow ~~~e_{}^{-x^{}_nx^{}_{n+1}\pi/4 }\,x^{}_n\,e_{}^{-x^{}_nx^{}_{n+1}\pi/4 }=x^{}_{n+1},
 $$
 from which 
 $$\mc H^{}_{n+1}=P\,\mc H^{}_n\,P^{-1}_{}=P^{n+1}_{}\mc H^{}_0\,P^{-n-1}_{},\quad \mc H^{}_0=e_{}^{iH' x_{1}x_{2}} \,e_{}^{iH^* x_{2N}x_{1}}=\mc H^{}_N,$$
 by periodicity. The wavefunctions
 $$\Psi_n=\mc H^{}_n\mc H^{}_{n-1}\cdots \mc H^{}_1\Psi^{}_0,$$
 obey the  recursion relation, ``{\em a Schr\"odinger equation for a discrete time variable!}",
 $$\Psi^{}_{n+1}=\mc H^{}_{n+1}\,\Psi^{}_n.$$
 
 The modified eigenfunction
 $$\Psi'_n=P^{-n}_{}\Psi^{}_n.$$
 also satisfies a recursion relation,
 $$\Psi_{n+1}'=\mc H^{}_0P^{-1}_{}\,\Psi'_n\equiv A\Psi'_n,$$
 but the shift operator, $\mc H^{}_0P^{-1}_{}$ does not depend on $n$. Nambu calls it $A$, but we call it $ F$ so as not to confuse with the matrix of the previous section. Then,
 
$$\Psi_N'=F^N_{}\Psi_0',~~~\longrightarrow ~~F^N_{}=1,$$
since  $F$ does not depend on $n$: the eigenvalues are roots of unity, $\lambda^N=1$. 
 
 \vskip .3cm
 \noindent  The eigenvalues are determined from the ``eigenoperator" equation that Nambu had previously used (see last section),  
 
 $$F\,X\,F^{-1}_{}=\lambda\,X,$$
 Its solution is expanded as a linear combination of $x_n$, 
 
 $$
 X=\sum_{n=1}^{2N} \alpha^{}_n\,x^{}_n.$$
 After  inserting this expansion in the eigenoperator equation,  the  expansion is written in terms of three coefficients, $a,b,c$,   
 $$
 X=\sum_{n=1}^{N-1}[a x^{}_{2n-1}+b x^{}_{2n}]\lambda_{}^{n-1}+c x^{}_{2N}+a x^{}_{2N-1}\lambda^{N-1}_{},$$
 reducing the eigenvalue operator equation  to  three coupled algebraic equations,
 \bean
\lambda^N{x}&=&\cosh2H^* \cosh2H'{x}-i\sinh2H^*\cosh2H'{y}+i\sinh2H' {z},\\
 \lambda{ z}&=&i\sinh2H^*{x}+\cosh 2H'{y},\\
  \lambda^{N-1}{ y}&=&-i\sinh2H'\cosh2H^*{x}-\sinh2H'\sinh2H^*{y}+\cosh2H'{z}. 
 \eean
 By eliminating the real variables $\bf x,y,z$ Nambu arrived at the consistency equation
 
 $$\lambda^{2N}+\sinh2H^*\sinh2H'(\lambda+\lambda^{-1}_{})-2\cosh2H^*\cosh2H'+\lambda^{-N}_{}=0,$$
 whose solution yields the eigenvalues. Setting $\lambda=e^{2\gamma}_{}$, it reduces to 

$$
\cosh2N\gamma=\sinh2H^*\sinh2H'(\lambda+\lambda^{-1}_{})-2\cosh2H^*\cosh2H'\cosh\gamma,
$$
to be solved for $\gamma$. He assumes $2\gamma=2\gamma_0+i\omega$, with $\omega=k\pi/N,\,k=1,2\dots,2N$. Comparing  the real and imaginary parts yields
$$
\pm\cosh 2\Gamma=\cosh2H^*\cosh2H'-\sinh2H^*\sinh2H'\cos\omega.$$
 Since the rhs is positive, it follows that 
 
 $$ \cosh 2\Gamma=\cosh2H^*\cosh2H'-\sinh2H^*\sinh2H'\cos\omega,\quad \omega=\frac{k\pi}{N},\quad k=1,2,\dots, N,$$
 which is Onsager's formula,  for large $N$, $N\gamma_0\rightarrow \Gamma$.
 \vskip .3cm
 \noindent{\bf Additional Remarks}
\vskip .2cm
\noindent  In the solution for the screw lattice,  Nambu emphasized a new mathematical method to solve eigenvalue problems which he had earlier used in his papers on ``{\em Third Quantization"}. He defined an ``eigenoperator" $X$  whose commutator with the operator  of interest is proportional to itself,
 
 $$[\,H,\,X\,]=\lambda X.$$
 Stated without proof are its  properties: 
 
 - $\lambda$ is the difference of two eigenvalues, $\lambda=E_n-E_m$.
 
 - $X$ transforms an eigenvector $\Psi_m$ of $H$ into another $\Psi_n$ with eigenvalue  $\lambda_n=E_m+\lambda$.
 
 - The product of the two eigenoperators $X_2X_1$ is again an eigen operator with eigenvalue $\lambda=\lambda_1+\lambda_2$, transforming an eigenvector to another one. 

- When $H$ has a simple structure, a general eigenoperator $X$ will be factorized into a product of  eigenoperators

$$X=X_1X_2\cdots X_k,\quad{\rm with~eigenvalues}~~e^\lambda_{}=e^{\lambda_{1}+\lambda_{2}+\dots \lambda_{k}}_{}.
$$

\section{Conclusions}
 Although Nambu's computation of the Ising model seems to be  a clever trick, the tremendous simplification  suggests that there must be conceptual advantages as well, possibly in the symmetries at the critical point, and perhaps connections to quantum codes.

\section{Acknowledgements}
This research was supported in part by the Department of Energy under Grant No. DE-SC0010296.

\end{document}